\documentclass{article}
\usepackage{csquotes}
\usepackage{graphicx}
\usepackage{amsmath}
\usepackage{amssymb}
\usepackage{setspace}
\usepackage{tabu}
\usepackage{biblatex}
\usepackage{caption}
\usepackage{subcaption}
\usepackage{inputenc}
\usepackage{geometry}
\begin{document}
\begin{figure}[!tbp]
  \hfill
  \begin{subfigure}[b]{0.48\textwidth}
    \includegraphics[width=\textwidth]{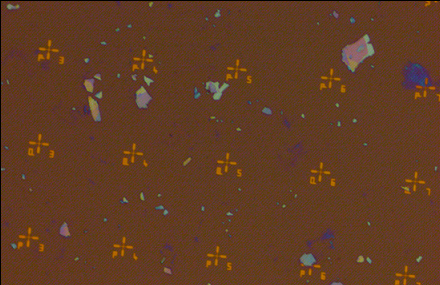}
    \caption{}
    \label{fig:f1b}
  \end{subfigure}
  \hfill
  \begin{subfigure}[b]{.48\textwidth}
    \includegraphics[width=\textwidth]{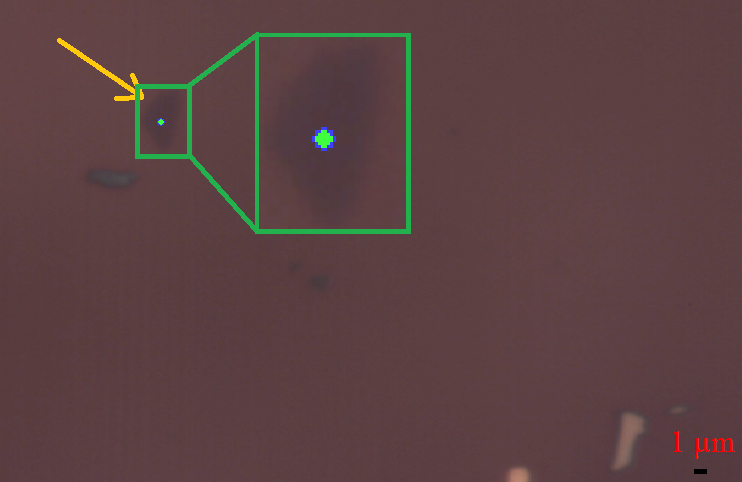}
    \caption{}
    \label{fig:f2a}
  \end{subfigure}
  \caption{(a) as deposited MoS\(_2\) flake using the scotch tape method with alignment marks (b) a few layer flake of MoS\(_2\)}
\end{figure}
\begin{figure}[!tbp]
  \begin{subfigure}[b]{0.48\textwidth}
    \includegraphics[width=\textwidth]{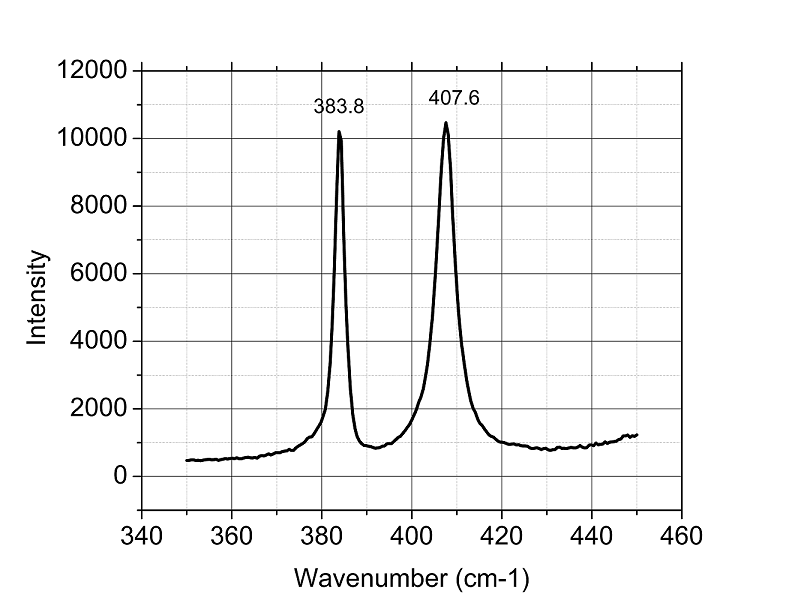}
    \caption{}
    \label{fig:f2b}
  \end{subfigure}
  \begin{subfigure}[b]{0.48\textwidth}
    \includegraphics[width=\textwidth]{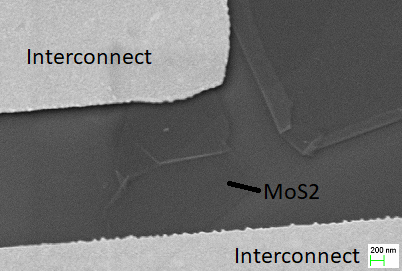}
    \caption{}
    \label{fig:f3b}
  \end{subfigure}
  \centering
  \begin{subfigure}[b]{.48\textwidth}
    \includegraphics[width=1\textwidth]{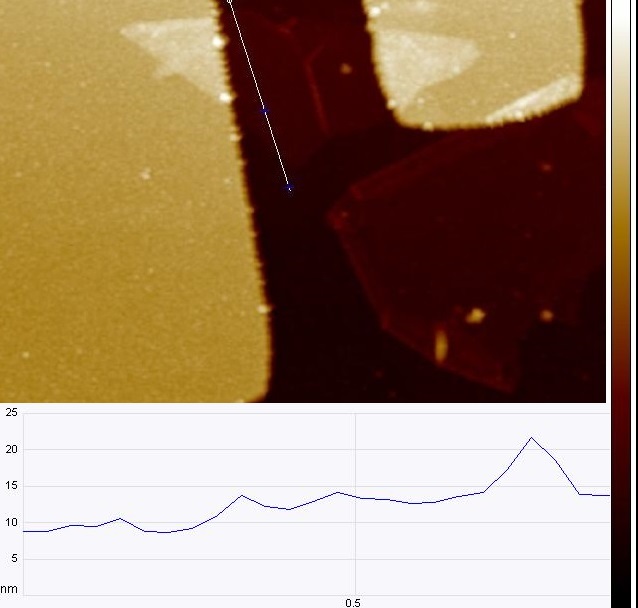}
    \caption{}
    \label{fig:f3a}
  \end{subfigure}
  \hfill
  \caption{(a) Raman spectroscopy data for a MoS\(_2\) flake. The peak to peak distance of 23.8 cm\(^{-1}\) indicates a few (4 or 5) layer flake (b) SEM image of a few layer MoS\(_2\) flake along with interconnects (c) AFM image of the same flake to confirm the MoS\(_2\) flake height}
\end{figure}
\begin{figure}[h!]
\centering
  \includegraphics[width=0.75\linewidth]{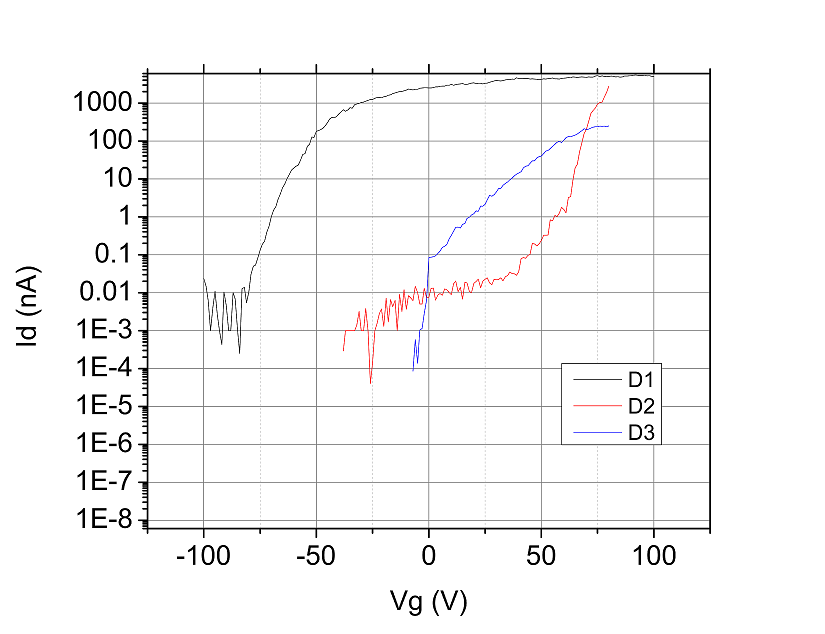}
  \caption{Semilog I\(_d\) vs V\(_g\) plot at V\(_d\) = 1 V for three devices fabricated. Considerable device to device variation exists due to different flake height i.e. number of MoS\(_2\) layers as well as the lateral size.}
  \label{fig:idvgsl}
\end{figure}
\begin{figure}[h!]
\centering
  \includegraphics[width=0.5\linewidth]{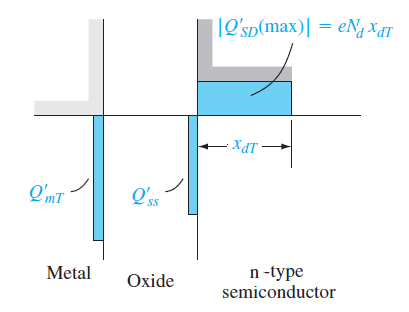}
  \caption{Charge distribution in a MOSCAP at threshold.}
  \label{fig:moscap}
\end{figure}
\begin{figure}[!tbp]
  \begin{subfigure}[b]{0.48\textwidth}
    \includegraphics[width=\textwidth]{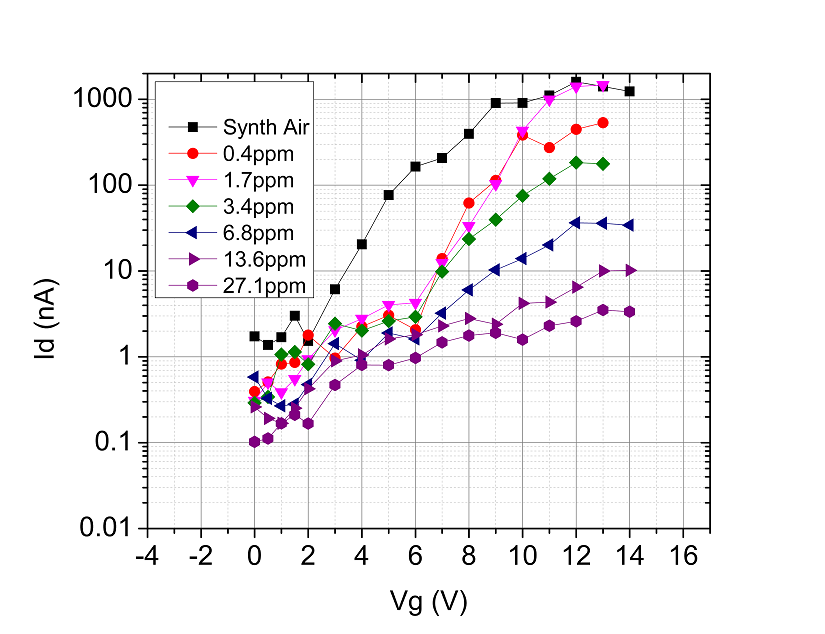}
    \caption{}
    \label{fig:f6a}
  \end{subfigure}
  \hfill
  \begin{subfigure}[b]{0.48\textwidth}
    \includegraphics[width=\textwidth]{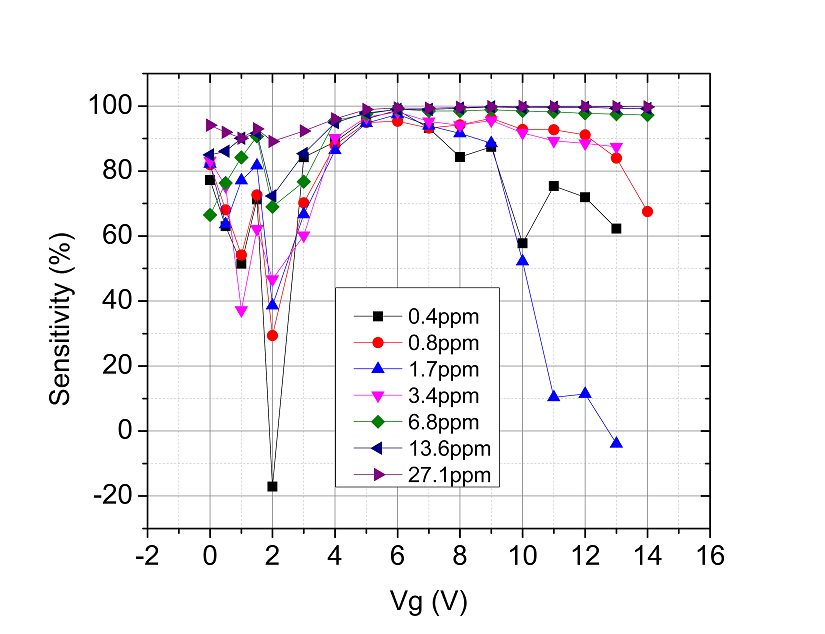}
    \caption{}
    \label{fig:f6b}
  \end{subfigure}
  \caption{(a) I\(_d\) vs V\(_g\) data at different acetone concentrations. As the acetone concentration increases, I\(_d\) for a given V\(_g\) decreases which can be explained by the gating effect of the electrons donated by acetone molecules (b) Calculated value of S (sensitivity) at different as V\(_g\) is sweeped at different acetone concentrations}
\end{figure}
\begin{figure}[!tbp]
  \begin{subfigure}[b]{0.48\textwidth}
    \includegraphics[width=\textwidth]{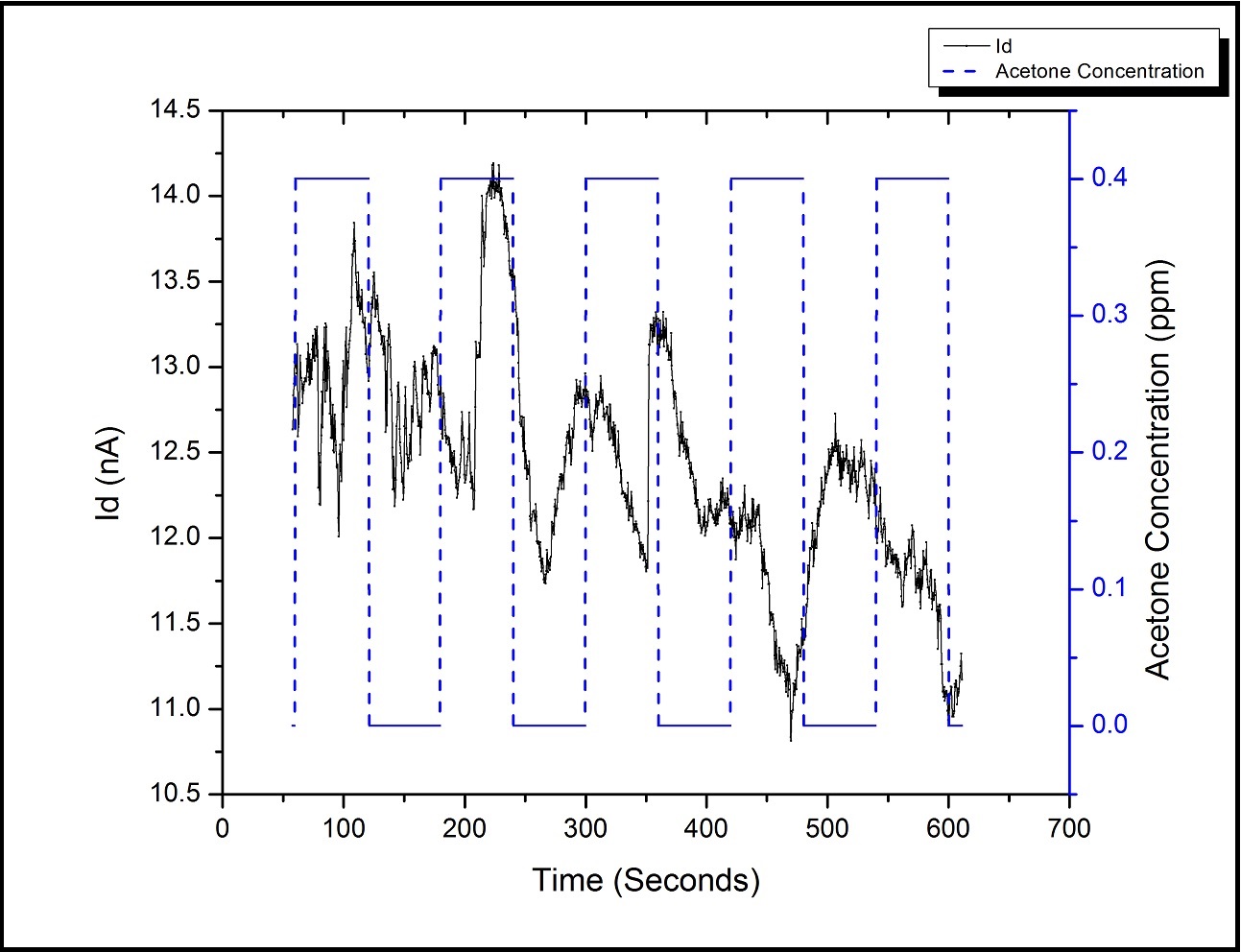}
    \caption{}
    \label{fig:f7a}
  \end{subfigure}
  \hfill
  \begin{subfigure}[b]{0.48\textwidth}
    \includegraphics[width=\textwidth]{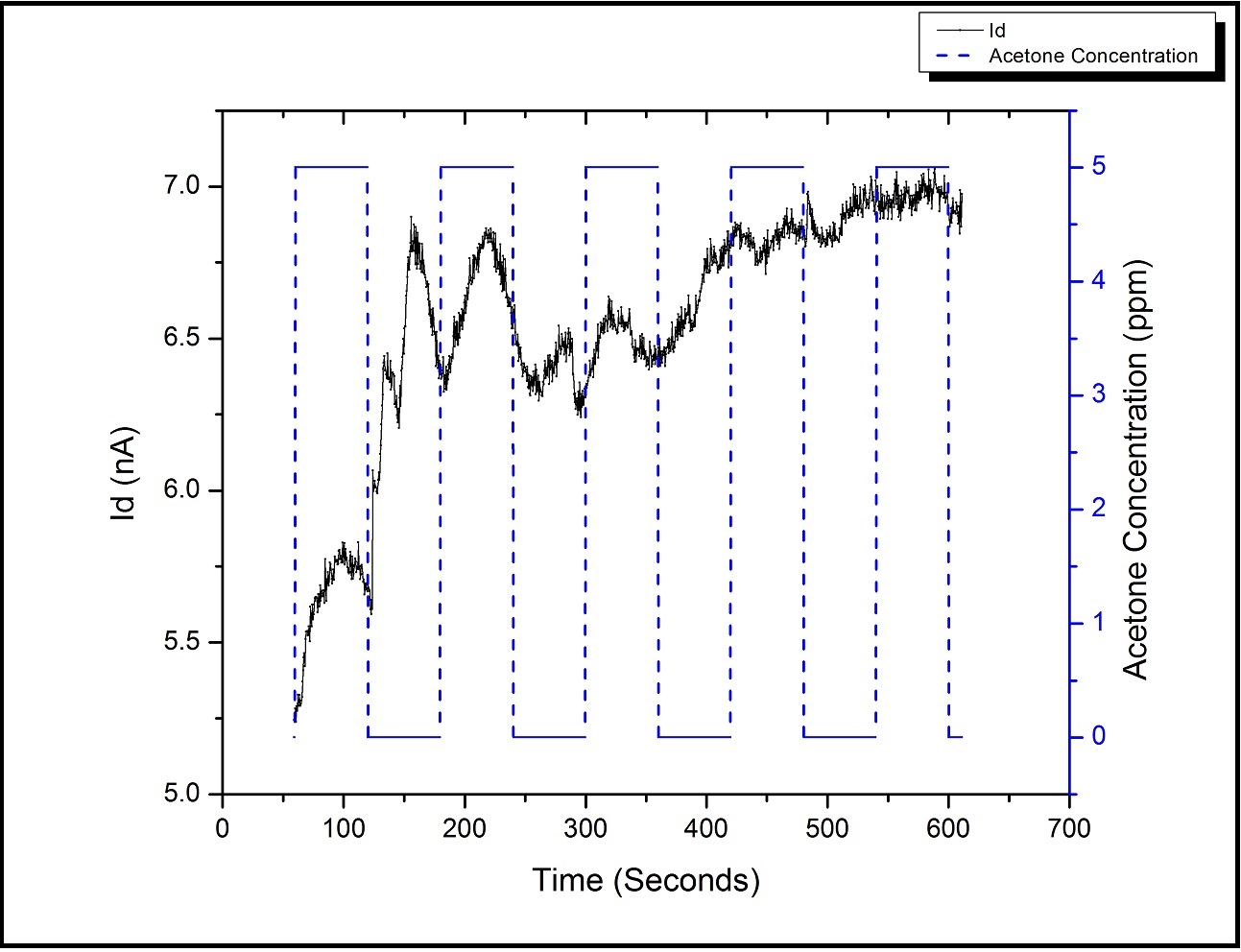}
    \caption{}
    \label{fig:f7b}
  \end{subfigure}
  \caption{Time response of device for 1 minute ON-OFF duty cycles at an acentone concentration of (a) 0.4 ppm (b) 5 ppm}
\end{figure}
\begin{figure}[h!]
\centering
  \includegraphics[width=0.5\linewidth]{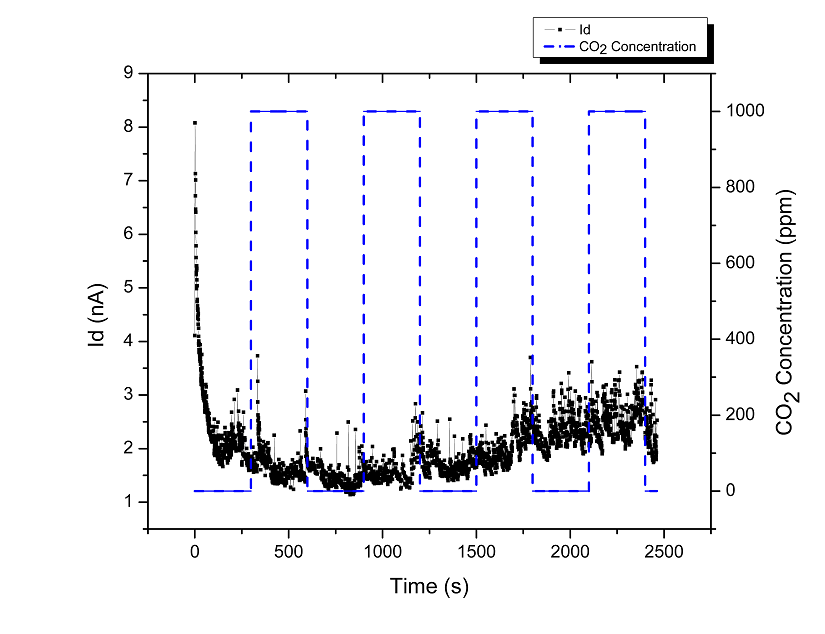}
  \caption{Selectivity test using CO\(_2\). The lack of functionalization for CO\(_2\) causes the lack of response in I\(_d\) to the presence of CO\(_2\). }
  \label{fig:f8}
\end{figure}

\textbf{ }
\newline
{\large \textbf{MoS\(_2\) Field Effect Transistors based sensors for low concentration acetone detection}}
\newline
\newline
Sumant Sarkar\(^1\), Alison Viegas\(^1\), and Srinivasa R. Raghavan\(^1\)
\newline
\(^1\)\textit{Indian Institute of Science, Bangalore, 560012, India}
\newline
Author to whom correspondence should be addressed; E-Mail: sraghavan@iisc.ac.in
\newline
\newline
\textit{Abstract:} The measurement of acetone in human breath, a known biomarker for diabetes, can act as an effective screening method for diabetes. While gas chromatography and mass spectroscopy based methods can detect gases in low concentration, they are costly, bulky, need complex sample preparation and are slow. Solid state devices such as Field Effect Transistors (FETs) can perform this faster and at a lower cost. In this work, we present the results of acetone gas sensing by molybdenum disulfide (MoS\(_2\)) based FETs. MoS\(_2\) channel was functionalized with polymethyl methacrylate (PMMA). When acetone molecules bind to PMMA, an electrostatic gating effect takes place and changes the electrical characteristics of the FET. More the number of acetone molecules that bind to the PMMA layer, the higher the shift in the drain current-voltage (I\(_d\)-V\(_g\)) curve of the FET. This shift is modeled as a change in the threshold voltage of the FET. The I\(_d\)-V\(_g\) curve was observed to shift downwards, as the acetone concentration increases. The limit of detection was observed to be 0.4 parts per million (ppm). This is the first sub-ppm concentration acetone detection at room temperature using a solid state device.
\newline
\newline
Keywords: molybdenum disulfide (MoS\(_2\)); acetone; gas sensor; diabetic; polymethyl methacrylate
\newline
\newline
\newline
\newline
\centerline{\sc \large INTRODUCTION}
\newline
\newline
\newline
\newline
There are 415 million people with diabetes in the world. By the year 2040, this number is expected to increase to 642 million. Approximately 1.5 to 5.0 million deaths each year resulted from diabetes\cite{1}. Diagnosis of diabetes requires a blood test and diagnostic infrastructure. Due to costs and other factors, in certain regions, two out of three cases remain undiagnosed. Chronic diabetes patients need to prick the skin to draw blood repeatedly. A point of care screening method that does not need to draw and process blood would be advantageous. Breath based screening methods which are reliable and can provide an immediate on-site results without needing any storage can be a solution to problems associated with the current mainstream method mentioned above.
\newline
If pathophysiological conditions such as diabetes cause a decrease in the glucose level in the body, ketone production in the body goes up. Acetoacetate (AcAc) and \(\beta\)-hydroxybutyrate are created in the liver from fatty acids. These compounds act as proxies for glucose. Acetone is a ketone formed by decarboxylation of AcAc and is excreted as waste via the lungs. This is the origin of higher acetone concentration in a diabetic person's breath. A diabetic person's breath has an acetone concentration of ranging from 1.7 ppm to 3.7 ppm. Breath of a healthy human being has acetone at a concentration of 0.8 ppm\cite{2}\cite{3}. Hence a sub-ppm level sensitivity is needed for diabetes screening. Human breath contains over two hundred kinds of volatile organic compounds (VOCs)\cite{4}. Hence the diagnostic method needs to be sufficiently selective for acetone.
\newline
Sub-ppm level gas sensing can be performed by Gas Chromatography (GC) and Mass Spectrometry (MS) methods\cite{5}. However, these methods are expensive, need bulky equipments, need substantial sample preparation and are not suitable for point-of-care diagnostics. Solid state device based methods are more suitable for such diagnosis as they are small and relatively inexpensive. Gas sensing has been performed using solid state devices such as transistors\cite{22-27}. Conventional silicon (Si) FET based sensors are inexpensive to make and are amenable to production automation. However, they suffer from poor sensitivity. 2D materials have a high surface area to volume ratio\cite{6}. Due to this, FETs based on 2D materials have an extremely high sensitivity and a low limit of detection (LOD). Graphene cannot be used to make FETs easily due to its lack of a bandgap. Transition metal dichalcogenides (TMDCs), such as MoS\(_2\) are suitable for this purpose since they have a bandgap. MoS\(_2\), which has a significant bandgap, has been known to provide high sensitivity, due to its high surface to volume ratio. Chemical Vapour Deposition (CVD) has promised mass production of MoS\(_2\)\cite{9}. Due to the low density of dangling bonds, flicker (white noise) is lower compared to Si based FETs. The use of MoS\(_2\) based FET to detect low concentration biomarker detection in liquid media has been demonstrated earlier\cite{6}\cite{7}\cite{14}. We present the results of gas sensing using FETs where the gas sensed is acetone and the FET channel is a two dimensional (2D), few atomic layers thin MoS\(_2\) flake. After fabricating the transistor, the drain current vs gate voltage (i.e. I\(_d\) vs V\(_g\)) performance is observed. The gate voltage at which the transistor turns ON (referred to as threshold voltage, V\(_{th}\)) is noted. The transistor current (I\(_d\)) is measured at V\(_{th}\), in the presence of synthetic air and it is compared to the current in the presence of acetone gas at different concentrations. 
\newline
\newline
\newline
\newline
\centerline{\sc \large MATERIALS AND METHODS}
\newline
\newline
\newline
\newline
A highly doped p++ silicon (Si) $<$100$>$ wafer with 300 nm silicon dioxide (SiO\(_2\)) grown on the polished side (Supplier: Graphene Supermarket) was used in the fabrication. The wafer was diced into 7 mm\(\times\)7 mm square shaped pieces using a laser dicing machine. Ellipsometer (M/S J. A. Woollam) was used to confirm the oxide thickness. The wafer piece was cleaned in standard RCA solution, followed by a 5 second dip in dilute hydrogen flouride (1:50). Four probe resistivity test was done to both sides of the sample to confirm oxide on one side (polished) and highly doped p++ Si on the other side. MoS\(_2\) flakes were first exfoliated on to a scotch tape. The wafer piece was heated at 120$^\circ$C for 2 minutes and then the scotch tape was used to transfer MoS\(_2\) flakes on to the wafer. Optical imaging was done to confirm the flakes and also identify potential few layer flakes. After this, the wafers were kept overnight in acetone to remove residual glue from tape. They were then washed with DI water and dried using N\(_2\) gun. Ammonium sulfide (Sigma Aldrich, 40-48 wt. \% in H\(_2\)O) was used for sulfur treatment. This step makes the contacts more ohmic and less schottky\cite{8}. After this, ebeam lithography (M/S Raith) and metallization was done to make Cr/Au (10/100nm) alignment marks. Raman spectroscopy (M/S LabRam) was used to identify MoS\(_2\) flakes that have a few layers. The alignment mark numbers were noted for those flakes. SEM (Zeiss Ultra 55) and AFM imaging were done to further characterize the MoS\(_2\) flakes. Graphic Database System (GDS) software was used for drawing the interconnects with the help of flake images and the alignment marks. This was followed by ebeam lithography and metallization steps to make source drain connections with Cr/Au (10/100 nm). Electrical characterization was done using an Agilent B1500 A Semiconductor Device Analyser to observe the transistor turn ON/OFF behavior and drain current characteristics as the gate voltage was varied. A functionalization layer with PMMA was applied by spin-coating at 12000 rpm. The PMMA thickness was measured using AFM and found to be 10 nm. A Printed Circuit Board (PCB) was fabricated using copper with pads for drain and source for each device on the sample and a common gate pad connected to the back side of the sample. The sample was placed on the PCB and stuck using silver paint. The device was then kept in a gas sensing setup with probe tips in contact with the gate, source and the drain. Source meters (Keithly 2450) was used in conjunction with Kick Start software to control and measure voltages and currents. I\(_d\) was measured at various V\(_g\) around the threshold voltage in: (a) ambient air (b) 100\% synthetic air supplied at 500 sccm (c) a mix of synthetic air and acetone gas at dfferent proportions (total sccm was always kept at 500). Based on the I\(_d\) - V\(_g\) data, sensitivity and other figures of merit were calculated.
\begin{figure}[h!]
\centering
  \includegraphics[width=0.75\linewidth]{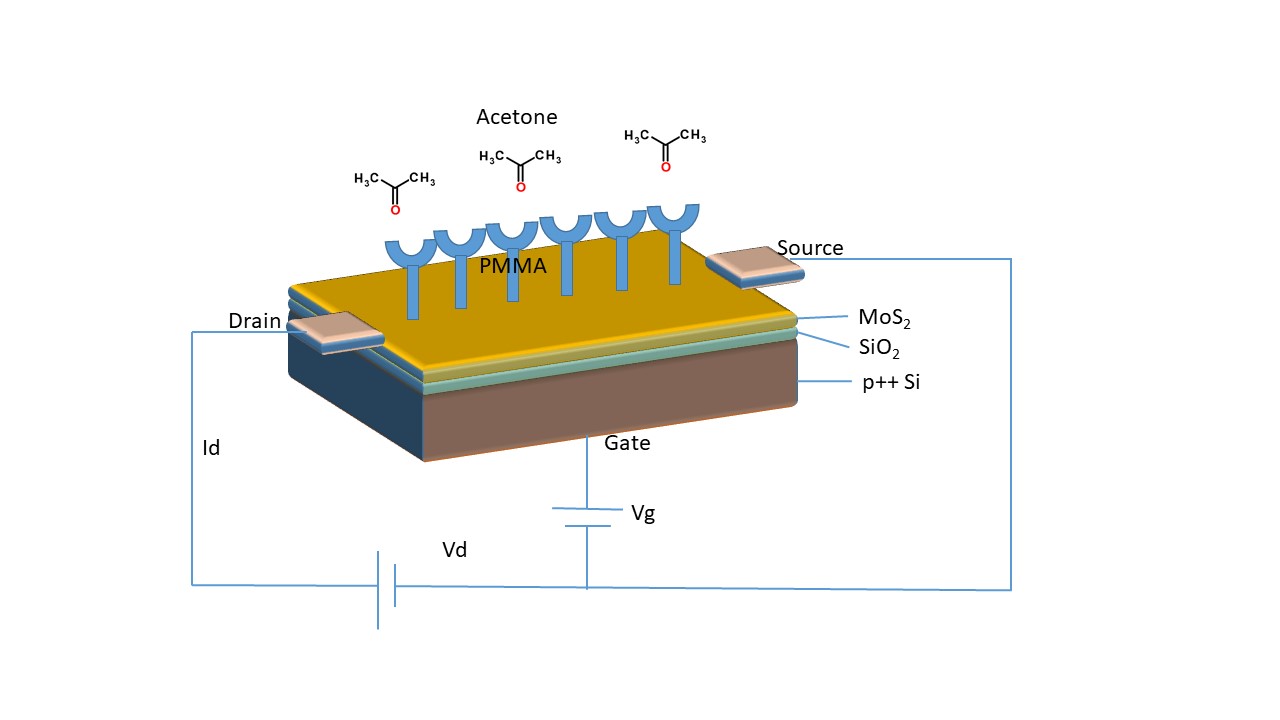}
  \caption{Schematic diagram of the device and circuit. The back gated structure allows unobstructed path to acetone molecules on to the PMMA layer}
  \label{fig:sddc}
\end{figure}
\newline
\newline
\newline
\newline
\newline
\newline
\newline
\centerline{\sc \large RESULTS AND DISCUSSION}
\newline
\newline
\newline
\newline
The oxide thickness on the Si wafer was found to be 287 nm. The electrical resistivity of the backside of the samples were found to be between 10.4 and 93.4 m\(\Omega/\Box\). This is sufficiently low for the purpose of back gating contacts. Optical images were taken after exfoliated MoS\(_2\) was transferred on to the wafer, using an optical microscope (Figure \ref{fig:f1b}). Raman spectroscopy peak to peak distance for MoS\(_2\) is 18.9 cm\(^{-1}\), 21.8 cm\(^{-1}\), 22.3 cm\(^{-1}\) and 27.0 cm\(^{-1}\) for 1 layer, 2 layer, 3 layer and bulk exfoliated MoS\(_{2}\), respectively\cite{9}\cite{10}\cite{11}\cite{12}\cite{13}. As Figure \ref{fig:f2b} shows, this distance was found to be 23.8 cm\(^{-1}\) indicating 4 or 5 layers of MoS\(_{2}\) in the flakes used to build FETs in this research.
Atomic Force Microscope (AFM) was used to obtain an accurate profile of the flakes and to measure the flake thickness more precisely (Figure \ref{fig:f3a}). 
Flakes ranging from 4 nm to 20 nm in thickness were found. Once the interconnects and contact pads were fabricated using ebeam lithography and metallization, electrical probe station (Agilent Device Analyzer B1500A) was used to perform the electrical characterization. Figure \ref{fig:idvgsl} shows the I\(_{d}\) vs V\(_{g}\) characteristics for three devices. The ON-OFF ratio was found to be of the order of 10\(^5\). The transconductance was measured to be between 150 nA/V and 200 nA/V. Subthreshold swing was calculated to be between 21000 mV/dec and 127 mV/dec. The flake lateral sizes were 3 \(\mu\)m \(\times\) 1 \(\mu\)m to 2 \(\mu\)m \(\times\) 2 \(\mu\)m.
Carrier mobility was computed using the expression 
\begin{align}
\mu = \left(\frac{\delta I_d}{\delta V_g} \right)\frac{L}{WC_{ox}V_d} 
\end{align}
where I\(_d\) is the drain current, V\(_g\) is the gate voltage, L is the channel length, W is the channel width, C\(_{ox}\) is the oxide capacitance per unit area and V\(_{d}\) is the drain voltage applied. Additionally, C\(_{ox}\) can be found with the formula: 
\begin{align}
C_{ox} = \frac{\epsilon}{t_{ox}}
\end{align}
Based on the data in our experiments, C\(_{ox}\) works out to 12 nF/cm\(^{2}\). Mobility for the devices works out to be between 12 and 36 cm\(^{2}\)/V.s which is comparable to the numbers in the literature\cite{9}.
\newline
\begin{figure}[!tbp]
  \begin{subfigure}[b]{0.48\textwidth}
    \includegraphics[width=\textwidth]{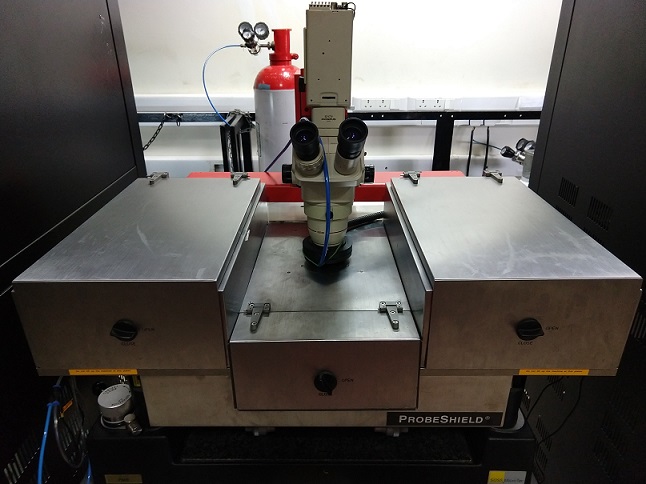}
    \caption{}
    \label{fig:f5a}
  \end{subfigure}
  \hfill
  \begin{subfigure}[b]{0.48\textwidth}
    \includegraphics[width=\textwidth]{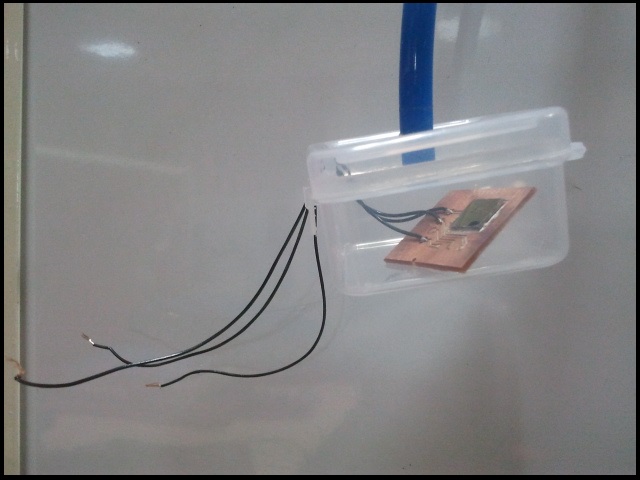}
    \caption{}
    \label{fig:f5e}
  \end{subfigure}
  \caption{Gas sensing setup used to test I\(_d\) sensitivity to acetone concentration at a fixed V\(_g\) and V\(_d\). It consists of the probe station chamber, source meters, MFCs with inlet and outlets and the MFC control software. Acetone gas  cylinder is (27.1 ppm) (b) Sample in a small chamber with an inlet for acetone and synthetic air}
\end{figure}
The devices chosen were the ones that had the threshold voltage in the 0 to 15 V range. This is where the sensitivity is the highest (Figure \ref{fig:f6b}). Gate voltage was sweeped in this range, keeping the V\(_{ds}\) fixed at 1 V. I\(_d\) was measured and changes were observed as the device was exposed to acetone of different concentrations. While calibrating the sensors, synthetic air was used as a baseline and as a diluent with a maximum flow of 500 sccm. The ratio of synthetic air and acetone was varied to get the required concentrations. As acetone concentration was increased from 0 to 27.1 ppm, its flow rate was increased from 0 to 500 sccm and synthetic air’s flow rate was decreased from 500 to 0 sccm, such that the total flow rate was always 500 sccm. This mechanism reduces extraneous factors and the changes in device current can be attributed to the acetone gas concentration alone.
\newline
In order to model the FET behavior at threshold, consider a MOSFET with a p-channel at the threshold inversion point. Figure \ref{fig:moscap} shows the charge distribution in a MOSFET at the threshold inversion point. The expression for the (ideal MOSFET under long channel assumption) threshold voltage is:
\begin{align}
V_{TP} = (-\vert Q'_{SD}(max)\vert - Q'_{SS})\frac{t_{ox}}{\epsilon_{ox}} + \phi_{ms} -2\phi_{fn}
\end{align}
where, \(V_{TP}\) is the threshold voltage, \(Q'_{SD}(max)\) is maximum space charge density per unit area of the depletion region, \(Q'_{SS}\) is the equivalent oxide charge, \(t_{ox}\) is the oxide thickness and \(\epsilon_{ox}\) is the oxide permittivity. One way to model the gas sensing is to calculate the expected change in this threshold voltage. As has been discussed in the literature\cite{28}, acetone molecules bind with PMMA chains. With the assumption of non-degenerative doping and that there is no material change in the MoS\(_2\) channel, we can state that there’s no change in last two terms in the equation for \(V_{TP}\). The charge transfer due to acetone binding can be modeled as a change in the depletion region space charge.
\newline
\begin{align}
\Delta V_{TP} = \frac{\Delta Q'}{C_{ox}} = \Delta Q' (\frac{t_{ox}}{\epsilon_{ox}})
\end{align}
where \(\Delta Q'\) is the charge transfer per unit area. The change in threshold voltage if a single electron transfer takes place in a 1 cm\(^2\) area, can be calculated from the above expression. It works out to 1.39 X 10\(^{-11}\) V or 13.9 pV. Considering a 1 um\(^2\) area, and assuming that n acetone molecules bind per PMMA site, the total shift in V\(_{th}\) is 3.953 X 10\(^{-16}\) X n Volts.
\newline
MoS\(_2\) is naturally slightly n-doped semiconductor. The electrons are the majority charge carries. Acetone, being a Lewis base, readily donates an electron. The charge transferred onto the PMMA layer causes a change in the channel conductivity as the electrons feel a repulsion from the electrons in the PMMA layer above. The carrier density in MoS\(_2\) is of the order 10\(^{12}\) – 10\(^{13}\) cm\(^{-3}\). The PMMA site density above is of the order of 10\(^{11}\) cm\(^{-2}\). Considering that multiple acetone molecules bind to a PMMA chain, the extent of electrostatic repulsion is significant. We expect a decrease in mobility, an increase in resistivity, a decrease in the transistor current when the device is exposed to acetone. Furthermore, this change will increase in magnitude as we increase the acetone concentration.
\newline
As can be observed from the plot in Figure \ref{fig:f6a}, there is indeed a decrease in the transistor current for given V\(_g\) when the device is exposed to acetone. The graph lines were no longer resolvable when the acetone concentration was reduced further down from 0.4 ppm. Additionally, the MFCs readings fluctuated substantially at this level making the measurement of acetone concentration unreliable. Furthermore, the extent of this change increases as the acetone concentration increases. From the raw data, we can calculate the sensitivity (S) with this expression:
\begin{align}
S = \frac{I_d^{air} - I_d^{acetone}}{I_d^{air}}
\end{align}
Using this formula, we calculated the sensitivity at different gate voltages for different acetone concentrations (Figure \ref{fig:f6b}). 
\newline
The sensitivities peaked around V\(_g\) = 6 V. Note that for a practical device, the sensitivities should be high. However, it is also desirable that there’s a large spread in sensitivities at different concentrations. It’s not good for a device, if the sensitivity is 100\% at all concentrations. Keeping this in mind, the ideal operating point for the device is Vg = 2 V to 2.5 V.
\newline
In the gas sensing measurements described above, the steady state values of I\(_d\) have been taken. Hence these experiments do not reveal anything on the dynamic response when the acetone concentration changes. Hence, a study was made on the time response of transistor currents as Acetone concentration was changed. One minute ON-OFF cycles were setup for two different acetone concentrations (0.4 ppm and 5 ppm) and the current variation was observed (Figure \ref{fig:f7a} and \ref{fig:f7b}).
\newline
In order to ascertain selectivity, the device was exposed to CO\(_2\) (Figure \ref{fig:f8}). CO\(_2\) was chosen due to its significant presence in human breath. No significant correlation was found between I\(_d\) and CO\(_2\) concentrations.
\newline
\clearpage
Kao et al. have summarized the literature on acetone sensing results\cite{5}. Ahn et al. have demonstrated acetone detection at room temperature. However the sensitivity was found to be 100 ppm\cite{18}. Kao et al. demonstrated a sensitivity of 0.4 ppm. However, this sensitivity was achieved at 200$^\circ$C. The current work has shown a room temperature sensitivity of 0.4 ppm.
\begin{center}
\begin{table}[t]
\caption{Comparison of acetone sensing approaches and results in the literature}
\begin{tabu} to 1\textwidth { ||X | X[2cm] | X | X|| }
\hline
 Material & Principle of operation & Lowest concentration detected (ppm)&Operating temperature\\
 \hline\hline
 In\(_2\)O\(_3\)\cite{15} & Resistence change in nanowire & 25& 400$^\circ$C \\ 
 \hline
 WO\(_3\)\cite{16} & Resistence change in nanoparticle & 0.2& 400$^\circ$C \\  
 \hline
 ZnO\cite{17} & Resistence change in thin film & 100& 200$^\circ$C\\    
 \hline
 ZnO+Ni+UV light\cite{18} & Resistence change in nanorods & 100& Room temperature\\    
 \hline
 LaFeO\(_3\)\cite{19} & Resistence change in thin film & 500& 275$^\circ$C\\    
 \hline
 TiO\(_2\)\cite{20} & Resistence change in thin film & 1& 500$^\circ$C\\    
 \hline
 GaN\cite{21} & Resistence change in thin film & 500& 350$^\circ$C\\    
 \hline
 InN\cite{5} & Resistence change in thin film & 0.4& 200$^\circ$C\\    
 \hline
 \textbf{MoS\(_2\) (this study)} & \textbf{Transconductance change FET} & \textbf{0.4}& \textbf{Room temperature}\\    
\hline
\end{tabu}
\end{table}
\end{center}
\centerline{\sc \large }
\centerline{\sc \large }
\centerline{\sc \large }
\centerline{\sc \large }
\centerline{\sc \large CONCLUSIONS}
\centerline{\sc \large }
\centerline{\sc \large }
\centerline{\sc \large }
\centerline{\sc \large }
Back gated FETs with a few layer MoS\(_2\) as channel were fabricated to make acetone sensors. The devices were characterized with AFM, Raman spectroscopy, SEM  and Electrical characterization. The FET was then used for gas sensing by studying the I\(_d\) change. The lower limit of concentration that could be detected was found to be 0.4 ppm. The ideal gate voltage where the sensitivity is high and its spread is maximum was found to be 2 V. To our knowledge, this is the first time detection of acetone gas at as low concentration as 0.4 ppm has been achieved using solid state devices at room temperature.
\newline
\newline
\newline
\newline
\centerline{\sc \large ACKNOWLEDGEMENTS}
\newline
\newline
\newline
\newline
This work was partly supported by Robert Bosch Centre for Cyber Physical Systems (RBCCPS) Research grant. Authors also acknowledge the help from Centre for Nanoscience and Engineering (CeNSE) at Indian Institute of Science. Authors also acknowledge the help from the staff at the CeNSE fabrication, characterization and other facilities.
\newline
\newline

\end{document}